\documentclass[a4paper,12pt, english]{article}
\usepackage[T1]{fontenc}
\usepackage{german}
\usepackage[latin1]{inputenc}
\usepackage{amsmath}
\usepackage{babel}
\usepackage{graphics}
\usepackage{amssymb}
\usepackage{amsfonts}
\usepackage{geometry}
\begin{document}

\title{Quasi-classical Approach to Bose Condensation in a Finite Potential
Well}

\author{L. A. Bányai, A. M. Bundaru\thanks{On leave of absence from Institute of Physics and Nuclear Engineering,
Bucharest }, and H. Haug\\
Institut für Theoretische Physik, J.W. Goethe-Universität,
  Frankfurt am Main }
\date{}
\maketitle

\begin{abstract}
We treat the problem of self-consistently interacting bosons in the
presence of a finite (but macroscopic) potential well within a quasi-classical
approximation for the normal component and the order parameter. We
solve the equilibrium problem and show, that actually condensation occurs in two steps.
One already at low densities with Bose condensation only in
the well and another one corresponding to the usual condensation in
bulk. The peak and width of the distribution of trapped particles in the well
display a distinct signature of the local condensation. A possible connection
to recent experiments with excitons is discussed.
\end{abstract}
\maketitle

\section{introduction}
 Since Bose-Einstein condensation (BEC) of trapped atoms was experimentally put in evidence
 \cite{Cornell,Ketterle} much theoretical work was devoted to the discussion of the various
 aspects of BEC in an external potential (see Refs. \cite{review, Pethick, PitString} for recent reviews).

In this paper we develop a systematic quasi-classical approach for the description of
 weakly interacting bosons with  condensate and non-condensate in equilibrium and non-equilibrium.
In order to describe  Bose-Einstein condensation of particles in a potential
well which contains many  closely spaced energy levels, a quasi-classical approach is
appropriate, because the detailed knowledge of the population of  higher
lying energy levels is neither of much interest and may cause an enormous
numerical effort for its determination.
Our attention is devoted mainly to finite potential wells, for which the trapping in the
 well is controlled by the bulk population.  
In a previous publication \cite{Schmitt} the evolution
of a non-interacting Bose gas coupled to a thermal reservoir towards
equilibrium in the presence
of a finite potential well with many energy levels was discussed. It was shown that
above a critical density the lowest state will be macroscopically
populated, i.e., the local density will go to infinity. In this paper
we take into account the repulsion between the bosons which eliminates
this unphysical behavior. We develop this theory within a quasi-classical
approach and a self-consistent treatment of the repulsion.
 We formulate
the equations describing the evolution towards equilibrium for the
normal component of the non-condensed particles, as well as for the condensate
order parameter. We discuss in
detail the exact equilibrium solution of these equations.
The resulting scenario is
a Bose condensation in two stages: At a first (rather low) critical
overall density a locally confined condensate appears in the well, its distribution 
broadens with increasing overall particle density and  
at a second (higher ) critical density one gets the usual bulk
condensate. The spatial distribution of the trapped particles changes abruptly
as the local condensation in the well occurs. The width of the distribution
displays as a function of the overall particle density a minimum in the form
of a cusp, while the slope of the peak changes discontinuously.

The described scenario may have a connection to recent experiments
\cite{Naka2} on excitons in a ${\rm Cu_2O}$ semiconductor under mechanical
stress, where a strong narrow luminescence from the center of the well was
observed.

\section{Quasi-classical description of a Bose system}

The Wigner function depending on the momentum \( \vec{p} \) and the
coordinate \( \vec{x} \) is defined as

\begin{equation}
\label{1}
f(\vec{p},\vec{x})=\int d\vec{y}e^{\imath \vec{p}\vec{y}\over \hbar }\langle \psi (\vec{x}+\frac{1}{2}\vec{y},t)^{+}
\psi (\vec{x}-\frac{1}{2}\vec{y},t)\rangle \, ,
\end{equation}
where \( \psi (\vec{x},t) \) is the second quantized wave function
and \( \langle \ldots \rangle  \) means averaging over a given ensemble.
This implies the normalization 

\begin{equation}
\label{1}
\int d\vec{x}\int \frac{d\vec{p}}{(2\pi \hbar )^{3}}f(\vec{p},\vec{x},t)=\langle N\rangle \, .
\end{equation}

All the averages of operators which are a sum of two operators 
\( O(\vec{x},\frac{\hbar }{\imath }\nabla )\equiv A(\vec{x})+B(\frac{\hbar }{\imath }\nabla ) \)
depending on  coordinate and momentum, respectively, may be
expressed as integrals \( \langle O(\vec{x},\frac{\hbar }{\imath }\nabla )\rangle
 =\int d\vec{x}\int \frac{d\vec{p}}{(2\pi \hbar )^{3}}f(\vec{p},\vec{x},t)O(\vec{x},\vec{p}) \).
In the quasi-classical limit \( f(\vec{p},\vec{x}) \) is real and
positive. For the case of an interaction with a thermal bath and slowly
varying potentials \cite{Kadanoff-Baym,Liboff} the following bosonic
kinetic equation holds

\begin{equation}
\label{Boltz}
\left( \frac{\partial }{\partial t}+\frac{1}{m}\vec{p}\,\nabla _{\vec{x}}-\nabla _{\vec{x}}U(\vec{x},t)
\nabla _{\vec{p}}\right) f(\vec{p},\vec{x},t)=-\int \frac{d\vec{p}\, '}{\hbar ^{3}}
\Bigl(   W(\vec{p},\vec{p}\, ')f(\vec{p},\vec{x},t)\left( 1+f(\vec{p}\, ',\vec{x},t)\right)
 -(\vec{p}\leftrightarrow \vec{p}\, ')\Bigr) \: ,
\end{equation}
where the positive transition rates \( W(\vec{p},\vec{p}\, ') \)
satisfy the detailed balance relation\begin{equation}
\label{balance}
W(\vec{p},\vec{p}\, ')=W(\vec{p}\, ',\vec{p})e^{\frac{\beta
    }{2m}(p^{2}-p'^{2})}\ ,
\end{equation}
and \( U(\vec{x},t) \) is a given potential. Generally speaking one
has an external potential \( U_{0}(\vec{x}) \), which we choose to
be time-independent (the given potential well) and an internal
one, due to the interaction between the bosons through a potential
\( v(\vec{x}) \) (where \( \vec{x} \) is the relative coordinate
of the two particles) related to the particle density \( n(\vec{x}) \)
(Hartree self-energy)\begin{equation}
\label{1}
U(\vec{x},t)\equiv U_{0}(\vec{x})+\int d\vec{x}\, 'v(\vec{x}-\vec{x}\, ')
n(\vec{x}\, ')=U_{0}(\vec{x})+\int d\vec{x}\, 'v(\vec{x}-\vec{x}\, ')
\int \frac{d\vec{p}}{(2\pi \hbar )^{3}}f(\vec{p},\vec{x}\, ',t)\: .
\end{equation}

The kinetic equation Eq.(\ref{Boltz}) has a single stable solution,
corresponding to equilibrium 

\begin{equation}
\label{Bose}
f_{eq}(\vec{p},\vec{x})=\frac{1}{e^{\beta \left(
      \frac{p^{2}}{2m}+U_{eq}(\vec{x})-\mu \right) }-1}\ ,
\end{equation}

where \( \mu  \) is the chemical potential and \( U_{eq}(\vec{x} \))
is determined by the self-consistency equation
\begin{equation}
\label{equil-s.c.}
U_{eq}(\vec{x})= U_{0}(\vec{x})+\int d\vec{x}\, 'v(\vec{x}-\vec{x}\, ')n(\vec x\;')= U_{0}(\vec{x})+\int d\vec{x}\, 'v(\vec{x}-\vec{x}\, ')
\int \frac{d\vec{p}}{(2\pi \hbar )^{3}}f_{eq}(\vec{p},\vec{x}\, ')\, .
\end{equation}

 The above described equations are  valid only in the absence of a condensation. As it
 is easy to understand, above a critical overall density \(\langle N\rangle\over V\)  the Bose distribution in 3D
 cannot provide
 a description of the whole particle density, and this is in conflict with the conservation
 of the particle number. It can be shown rigorously \cite{condboltz} that    
 for super-critical conditions,  where a condensation
occurs,  Eq.(\ref{Boltz}) has no solutions with ordinary functions but only with
 distributions.

 An appropriate approach to avoid this problem is to consider a two-component treatment where
\(f(\vec{p},\vec{x},t) \) describes only the non-condensate and a condensate is taken into account explicitly.

The existence of such a phase transition with spontaneous symmetry breaking in the presence  of a potential
 is of course  an idealization, which admits that the distance between the energy
 levels is negligible. This admition is implicitely made by the quasiclassical approximation.
A  discussion of the meaning of BEC in systems with discrete spectrum is given in Ref.\cite{condvol}. 
    
In the presence of a condensate the s.c. potential is then given by the total density (non-condensate and condensate)
 \begin{equation}
\label{densit}
n(\vec{x},t)= \int\frac{d\vec{p}}{(2\pi \hbar )^{3}}f(\vec{p},\vec{x},t)+|\langle \psi (\vec{x}\,t)\rangle |^{2}\, , 
\end{equation}
where \(\langle \psi (\vec{x}\,t)\rangle\) is the condensate amplitude  (average of the second quantized wave function)
 and thus the two components are coupled.  A further coupling will result from irreversible
 transitions from and to the condensate.

In order to include  the condensate  one needs an equation for its amplitude.
For the mathematical  consistency of the whole  model, however, the quasi-classical
 approximation  for the condensate  also has to be formulated.
One may start from the Schrödinger equation with the self-consistent (s.c.) potential \( U({\vec x}, t)\) 
for the   condensate ampitude  
 \begin{equation}
\label{1}
\imath \hbar \frac{\partial }{\partial t}\langle \psi (\vec{x},t)\rangle =
\left( -\frac{\hbar ^{2}}{2m}\nabla ^{2}+U(\vec{x},t)\right) \langle \psi (\vec{x},t)\rangle \: .
\end{equation}
Neglecting the contribution of the non-condensate to the s.c. potential, this equation 
coincides with the Gross-Pitaevskii equation, but we do not neglect this contribution.

One has to recall here, that the quasi-classical limit means  \( \hbar \to 0 \)  and therefore
in this limit the kinetic energy term disappears (but as usual the  \( \hbar  \) from the time derivation 
survives)  and we get
\begin{equation}
\label{Pitaevskii}
\imath \hbar \frac{\partial }{\partial t}\langle \psi (\vec{x},t)\rangle =U(\vec{x},t) \langle \psi (\vec{x},t)\rangle \: . 
\end{equation}

Next, as already mentioned, we have to include the transitions between the normal component and condensate
 in the Eqs.(\ref{Boltz},\ref{Pitaevskii}).

In the kinetic equation Eq.(\ref{Boltz}) we have to add  the scattering rate
 due to the transitions from the state \(\vec p\) to the \({\vec p}=0\)
 condensate 
\begin{equation}
\label{colterm1a}
- W(\vec{p},0)f(\vec{p},\vec{x},t)|\langle \psi (\vec{x},t)\rangle |^{2} \: .
\end{equation}
The initial state has to be populated, hence the factor \(f(\vec{p},\vec{x},t)\).
The final state factor contains the term due to stimulated transitions 
\(\propto |\langle \psi (\vec{x},t)\rangle |^{2}\), because the term
describing spontaneous transition is lost 
in the thermodynamic limit (see also Ref. \cite{condboltz}). Similarly the
scattering rate due to 
transitions from the condensate to a  state \(\vec p\) is given by 
\begin{equation}
\label{colterm1b}
W(0,\vec{p})(1+f(\vec{p},\vec{x},t)) |\langle \psi (\vec{x},t)\rangle |^{2} \: .
\end{equation}
Formally, these terms result by adding to the distribution function 
\( f(\vec{p}\, ',\vec{x},t) \)
in the collision term of Eq.(\ref{Boltz}) a condensate term
 \( (2\pi \hbar )^{3}\delta (\vec{p}\, ')|\langle \psi (\vec{x},t)\rangle
 |^{2}\ .\)

The gain and loss rates for the order parameter amplitude have to
 be calculated by taking the initial and final state factors for the
condensate to be one
\begin{equation}
\frac{\imath \hbar }{2}\langle \psi (\vec{x},t)\rangle \int \frac{d\vec{p}}{(2\pi \hbar )^{3}}
\Bigl( W(\vec{p},0)f(\vec{p},\vec{x},t)-W(0,\vec{p})(1+f(\vec{p},\vec{x},t))
\Bigr) \: .
\end{equation}
Such collision terms have been obtained \cite{condboltz} in the frame
of the equation of motion method for a system of bosons weakly interacting
with a thermostat, decoupling higher order correlations and taking
the long-time Markovian limit (see also \cite{condvol}). A more involved approach, within the
frame of quantum kinetic, taking into account time-dependent energy
renormalizations due to boson-boson interaction was described in Ref.
\cite{kinquantbos}. Neglecting these energy corrections and taking the
 long-time limit one recovers the previous results.

The resulting system of coupled kinetic equations which describe the quasi-classical
evolution of  interacting Bosons within the self-consistent approximation
as

\[
\left( \frac{\partial }{\partial t}+\frac{1}{m}\vec{p}\,\nabla _{\vec{x}}-
\nabla _{\vec{x}}U(\vec{x},t
)\nabla _{\vec{p}}\right) f(\vec{p},\vec{x},t)=
-\int \frac{d\vec{p}\, '}{(2\pi \hbar )^{3}}
\Bigl( W(\vec{p},\vec{p}\, ')f(\vec{p},\vec{x},t)
\left( 1+f(\vec{p}\, ',\vec{x},t)\right) -(\vec{p}\leftrightarrow \vec{p}\, ')\Bigr) \]

\begin{equation}
\label{Boltz-cond}
-\Bigl( W(\vec{p},0)f(\vec{p},\vec{x},t)-W(0,\vec{p})(1+f(\vec{p},\vec{x},t))
\Bigr)
 |\langle \psi (\vec{x},t)\rangle |^{2}\ ,
\end{equation}
together with
\begin{equation}
\label{ec-condens}
\imath \hbar \frac{\partial }{\partial t}\langle \psi (\vec{x},t)\rangle =U(\vec{x},t)\langle \psi (\vec{x},t)\rangle +\frac{\imath \hbar }{2}\langle \psi (\vec{x},t)\rangle
 \int \frac{d\vec{p}}{(2\pi \hbar )^{3}}\Bigl( W(\vec{p},0)f(\vec{p},\vec{x},t)-W(0,\vec{p})(1+f(\vec{p},\vec{x},t))\Big ) \, ,
\end{equation}

and of course with the self-consistency condition
 \begin{equation}
\label{pot-s.c.}
U(\vec{x},t)=U_{0}(\vec{x})+\int d\vec{x}\, 'v(\vec{x}-\vec{x}\, ')\left( \int \frac{d\vec{p}}{(2\pi \hbar )^{3}}f(\vec{p},\vec{x}\, ',t)+|\langle \psi (\vec{x}\, ',t)\rangle |^{2}\right) \: .
\end{equation}

The equation for the condensate density \( |\langle \psi (\vec{x})\rangle | \)
emerges as

\begin{equation}
\label{denscon}
\frac{\partial }{\partial t}|\langle \psi (\vec{x},t)\rangle |^{2}=|\langle \psi (\vec{x},t)\rangle |^{2}\int \frac{d\vec{p}}{(2\pi \hbar )^{3}}\Bigl( W(\vec{p},0)f(\vec{p},\vec{x},t)-W(0,\vec{p})(1+f(\vec{p},\vec{x},t))\Bigr ) \, .
\end{equation}

One sees that the equations conserve the total average number of bosons

\begin{equation}
\label{number}
\int d\vec{x}|\langle \psi (\vec{x},t)\rangle |^{2}+\int d\vec{x}\int \frac{d\vec{p}}{(2\pi \hbar )^{3}}f(\vec{p},\vec{x},t)=\langle N\rangle \, .
\end{equation}

These equations describe an irreversible evolution.  Above a certain overall particle density 
(at a given thermostat temperature)  any arbitrarily small, but non-vanishing
initial condensate in all points \( \vec{x} \) will evolve to its
finite  stable equilibrium (non-homogeneous) value, while below this density  it will disappear. 

Because the \( \hbar \to 0 \) limit for the Schrödinger equation
is equivalent to the infinite mass limit, in this quasi-classical approximation
the condensate does not propagate, but will be created or destroyed
locally.

A theory, in which the quasi-classical approximation is used only for the non-condensate,
  runs into mathematical inconsistencies.

The theory we described takes into account the effect of the particle repulsion  for the spatial distribution only. Anomalous pair correlations giving rise to a modified quasi-particle (Bogolyubov-) spectrum were ignored. 

 As for any quasi-classical approach we expect the  validity of the above described theory for
 slowly varying potentials where the distance between the energy levels is very small. A detailed
 discussion of these conditions is given in the last section, where the theory is tentatively applied
 to interpretation of experimental data.
    
For a local interaction \( v(\vec{x})=w\delta (\vec{x}) \) the self-consistency
equation becomes also local\begin{equation}
\label{pot.sc.-loc}
U(\vec{x},t)=U_{0}(\vec{x})+w\left( |\langle \psi (\vec{x},t)\rangle |^{2}+\int \frac{d\vec{p}}{(2\pi \hbar )^{3}}f(\vec{p},\vec{x},t)\right) \, ,
\end{equation}
which facilitates its treatment. In what follows we shall restrict
the discussion to the equilibrium problem with such a local interaction.

\section{Bose condensation}

We discuss here the equilibrium solution of Eqs.(\ref{Boltz-cond}-\ref{pot-s.c.})
considering a local interaction with \( w>0 \) (repulsion) at a given
temperature for various overall densities \( \bar{n}\equiv \frac{\langle N\rangle }{V} \)
. Here the thermodynamic limit is implicitly understood. 

In the absence of an equilibrium condensate \( |\langle \psi (\vec{x})\rangle _{eq}|^{2}=0 \),
the equilibrium solution for the normal component is the Bose distribution
Eq.(\ref{Bose}). Above a certain overall density \( \bar{n} \) (see
Eq.(\ref{number}) divided by the volume \(V\)  ) a local condensate appears.
 The normal component
should be further distributed according to the Bose distribution,
which is a stationary solution of the l.h.s. of the kinetic equation
Eq.(\ref{Boltz-cond}) and of the part of the collision term which
does not involve the condensate. The equilibrium condensate must be
such as to cancel the additional collision term

\begin{equation}
|\langle \psi (\vec{x},t)\rangle |^{2}\int \frac{d\vec{p}}{(2\pi \hbar )^{3}}
\Bigl ( W(\vec{p},0)f(\vec{p},\vec{x},t)-W(0,\vec{p})(1+f(\vec{p},\vec{x},t))\Bigr ) 
\end{equation}
of both equations Eq.(\ref{Boltz-cond}) and Eq.(\ref{denscon}).
Inserting the Bose function, the equilibrium condensate density \( n_{0}(\vec{x}) \)
together with the detailed balance relation Eq. (\ref{balance}),
this term becomes
\begin{equation}
n_{0}(\vec{x})\int \frac{d\vec{p}}{(2\pi \hbar )^{3}}W(\vec{p},0)f_{eq}(\vec{p},\vec{x})\left( 1-e^{\beta (U_{eq}(\vec{x})-\mu )}\right) \: .
\end{equation}
As a result,  the equilibrium condensate density \( n_{0}(\vec{x}) \)
may be different from zero only, where \( U_{eq}(\vec{x})-\mu =0 \).
Due to  the positivity of \( f_{eq}(\vec{p},\vec{x}) \), the chemical
potential always must lie below the minimum of the potential \( \mu \leq min U_{eq} \). Therefore the condensate can exist only in the minimum of the equilibrium
self-consistent potential. This again corresponds to the behavior of particles
with an infinite mass. 

The order parameter itself cannot be constant in time, but according
to Eq.(\ref{ec-condens}) oscillates as \( \langle \psi (\vec{x},t)\rangle _{eq}=e^{-\imath \mu t}\sqrt{n_{0}(\vec{x})}\)
and it is actually determined only up to an arbitrary coordinate-dependent
phase factor, which is determined from the initial condition. In general
the order parameter in equilibrium always oscillates with the chemical
potential because the particle number operator does not commute with
the macro-canonical density matrix in the case of the spontaneous symmetry
breaking (see \cite{kinquantbos}). 

Without repulsion, the chemical potential can increase with the overall
density up to the minimum of the potential \( U_{0}(\vec{x}) \).
Further increase of the overall density \( \bar{n} \) is possible
only, if the local condensate density \( n_{0}(\vec{x}) \) itself
increases as the volume. This means that the assumed slow spatial
variation which was required in the derivation of the quasi-classical
description, breaks down. However, we shall see that the situation
changes substantially by the existence of a repulsion between the
bosons, allowing a quasi-classical description of BEC.

Let us discuss, what happens in the presence of a repulsion when one
increases the chemical potential starting form below the minimum of
the external potential \( U_{0}(\vec{x}) \). This is of course equivalent
to increasing the overall density \( \bar{n} \). 

The self-consistency condition for equilibrium with a local interaction
is 

\begin{equation}
\label{1}
U_{eq}(\vec{x})=U_{0}(\vec{x})+w\left( n_{0}(\vec{x})+\int \frac{d\vec{p}}{(2\pi \hbar )^{3}}f_{eq}(\vec{p},\vec{x})\right) \: .
\end{equation}
This equation has been extensively used in the equilibrium theory of Bose-Einstein condensation of atoms trapped in an oscillator potential
( see Ref.\cite{review} and references therein), where similar considerations
apply, although  the nature of the trapping kinetics is different.

We choose an attractive external potential \( U_{0}(\vec{x})\leq 0 \)
which vanishes outside a finite domain around \( \vec{x}=0 \), and
has a minimum in \( \vec{x}=0 \) . One may expect that the potential
\( U_{eq}(\vec{x}) \) also will have a point of minimum or a minimum
surface and will be constant outside the range of \( U_{0} \) . Generally
speaking \( minU_{eq}>minU_{0} \) , since the repulsion opposes the
attractive external potential .\\

For a repulsive interaction the increasing density increases also
the minimum of \( U_{eq}(\vec{x}) \) and eventually flattens the
bottom of the potential. The number of locally condensed particles
will still remain finite and therefore does not contribute to the
overall density. Only for an overall density \( \bar{n} \) greater
than the critical density \begin{equation}
\label{1}
n_{c}=\int \frac{d\vec{p}}{(2\pi \hbar )^{3}}\frac{1}{e^{\beta \frac{p^{2}}{2m}}-1}
\end{equation}
an overall condensate appears. 

Because outside the range of \( U_{0}(\vec{x}) \) both the local
density and the self-consistent potential are constant, we have for \( \bar{n}\leq n_{c} \)
the following equation for the determination of the chemical potential 

\begin{equation}
\label{1}
\int \frac{d\vec{p}}{(2\pi \hbar )^{3}}\frac{1}{e^{\beta \left( \frac{p^{2}}{2m}+w\bar{n}-\mu \right) }-1}=\bar{n}\, .
\end{equation}
This equation has a solution only for \( \mu \leq wn_{c} \) . Above this
value an overall condensate has to appear. \\

For simplicity let us consider a spherically symmetric situation (\( U_{0}(r) \)
with \( U_{0}(r)=0 \) for \( r>R \) and a negative minimum in \( r=0 \)
). We consider the case where \( \mu  \) is above the minimum of
\( U_{0} \), but below \( wn_{c} \) . Let us look at the shape of
\( U_{eq}(r) \) coming from large \( r \) . Here one has a spatially
constant self-consistent potential \begin{equation}
\label{1}
U_{eq}(r)=w\bar{n};\; (r>R)\: .
\end{equation}

For \( r<R \) one gets a radius-dependent density and  the self-consistency equation 

\begin{equation}
\label{1}
U_{eq}(r)=U_{0}(r)+w\int \frac{d\vec{p}}{(2\pi \hbar )^{3}}\frac{1}{e^{\beta \left( \frac{p^{2}}{2m}+U_{eq}(r)-\mu \right) }-1};\: (r_{0}<r<R)
\end{equation}

for \( U_{eq}(\vec{x})-\mu >0 \) giving rise to a negative \( U_{eq}(r) \)
varying monotonously. At a certain \( r=r_{0} \) one may get \( U_{eq}(r_{0})=\mu  \)
and the integral achieves its maximal value \( n_{c} \) . Due to
the condition \( U_{eq}(\vec{x})-\mu \geq 0 \) all the points \( r<r_{0} \)
must belong to the minimum of a monotonous \( U_{eq}(r) \) and a
condensate \( n_{0}(r) \) must emerge 

\begin{equation}
\label{1}
U_{eq}(r)=U_{0}(r)+w(n_{c}+n_{0}(r));\: (r<r_{0})
\end{equation}
in order to ensure that \( U_{eq}(r) \) remains constant for \( r<r_{0} \)
, i.e. , 

\begin{equation}
\label{1}
U_{eq}(r)=U_{eq}(r_{0}+0)=\mu \; (r<r_{0})\: ,
\end{equation}
 implying 

\begin{equation}
\label{1}
n_{0}(r)=\frac{1}{w}\left( U_{0}(r_{0})-U_{0}(r)\right) \; (r<r_{0})\: .
\end{equation}

When the overall density \( \bar{n} \) reaches the critical value
\( n_{c} \) , i.e. \( \mu =wn_{c} \) , the self-consistent potential is completely
flat, \( r_{0} \) approaches \( R \), while the local condensate
density approaches \( -\frac{1}{w}U_{0}(r) \) .

For overall densities above the critical one, one gets an overall
condensate, while the local condensate density is \( \bar{n}-n_{c}-\frac{1}{w}U_{0}(r) \).

This scenario is complete and self-consistent!

However, local condensation shows non-analyticities
only  of local entities, but does not affect the analyticity of the total thermodynamic entities.
Therefore according to the usual terminology it does not define a phase transition.

We expect the  following scenario in a self-consistent but quantum mechanical picture:
In the presence of a repulsive interaction between the particles the chemical potential approaches
 the lowest s.c. level and pushes it up without actually touching it until the bound states
 disappear completely and the true bulk phase transition occurs. In this way a finite local density
 appears first, which can be higher than the critical density of the bulk.

\section{Numerical illustration}

For the following we choose a spherically symmetric, attractive (\( v<0 \)
) external potential

\begin{equation}
\label{1}
U_{0}(r)=v\left( 1-\left( \frac{r}{R}\right) ^{2}\right) \: .
\end{equation}

For further discussions it is convenient to measure densities in units
of the critical density \( n_{c}, \) the energies in units of \( k_{B}T \)
(\( {\cal U}\equiv \beta U_{eq} \), \( {\cal U}_{0}\equiv \beta U_{0} \)
, \( {\cal M}\equiv \beta \mu  \) , \( {\cal V}\equiv \beta v \), 
the radius in units of \( R \) (\( \rho \equiv r/R \) ), and use
the dimensionless interaction constant \( {\cal W}\equiv \beta wn_{c} \).

Then \begin{equation}
\label{1}
{\cal U}_{0}(\rho )={\cal V}\left( 1-\rho ^{2}\right) \: ({\cal V}<0)
\end{equation}

and one has to solve the equation \begin{equation}
\label{transceq}
{\cal U}(\rho )={\cal U}_{0}(\rho )+{\cal W}\frac{1}{\int _{0}^{\infty }ds\frac{\sqrt{s}}{e^{s}-1}}\int _{0}^{\infty }ds\frac{\sqrt{s}}{e^{s+{\cal U(\rho )}-{\cal M}}-1}
\end{equation}
 in every point \( \rho _{0}<\rho <1 \) on order to reconstruct the
condensate density as describe above. Obviously the radius \( \rho _{0}=\frac{r_{0}}{R} \)
for which the self-consistent potential touches the chemical potential is defined
by the solution of the equation

\begin{equation}
{\cal M}={\cal V}\left( 1-\rho _{0}^{2}\right) +{\cal W}
\end{equation}

 which exists in the interval \( 0<\rho _{0}<1 \) only for \( {\cal M}>{\cal V}+{\cal W} \)
.

Since the transcendental equation Eq.(\ref{transceq}) is local, for
a numerical solution it is convenient to solve it in favor of \( {\cal U}_{0} \)
at a given \( {\cal U} \), thus performing a simple integration.
The association to a certain radius is given by the explicit definition
of \( {\cal U}_{0}(\rho ) \).

We illustrate the above described scenario with some
numerical examples. We choose \( {\cal V}=-5 \) and \( {\cal W}=0.05 \)
and discuss later, what happens for stronger coupling interaction
potentials. 

\begin{figure}
{\centering \resizebox*{12cm}{8cm}{\includegraphics{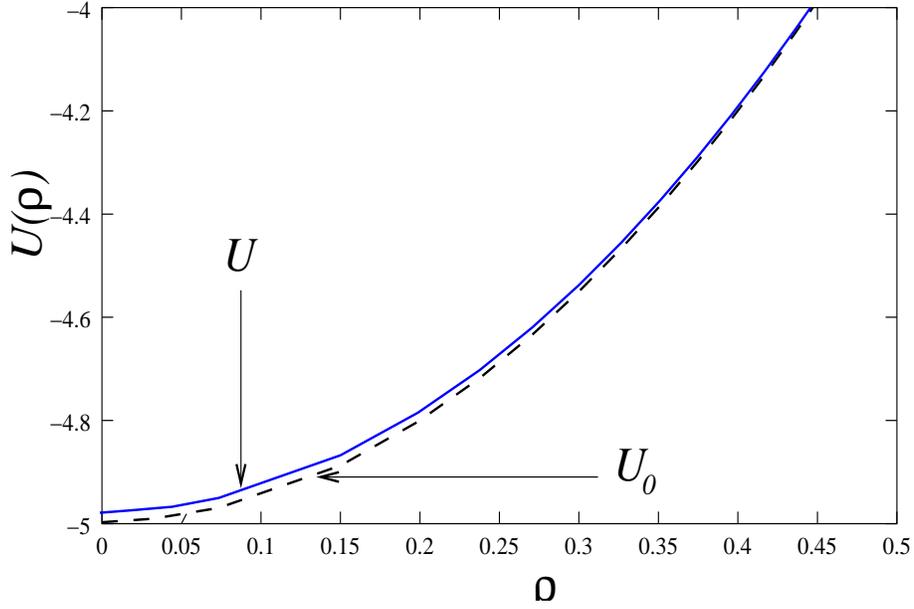}} \par}

\caption{External potential \protect\( {\cal U}_{0}(\rho )\protect \) (dashed
line) and self-consistent potential \protect\( {\cal U}(\rho )\protect \) (full
line) for a sub-critical condition \protect\( {\cal M}=-5.2\protect \)
(\protect\( \bar{n}=0.0021n_{c}\protect \) ) with a potential depth
\label{pot-1} \protect\( {\cal V}=-5\protect \), and an interaction constant
\protect\( {\cal W}=0.05\protect \).}
\end{figure}

\begin{figure}
{\centering \resizebox*{12cm}{8cm}{\includegraphics{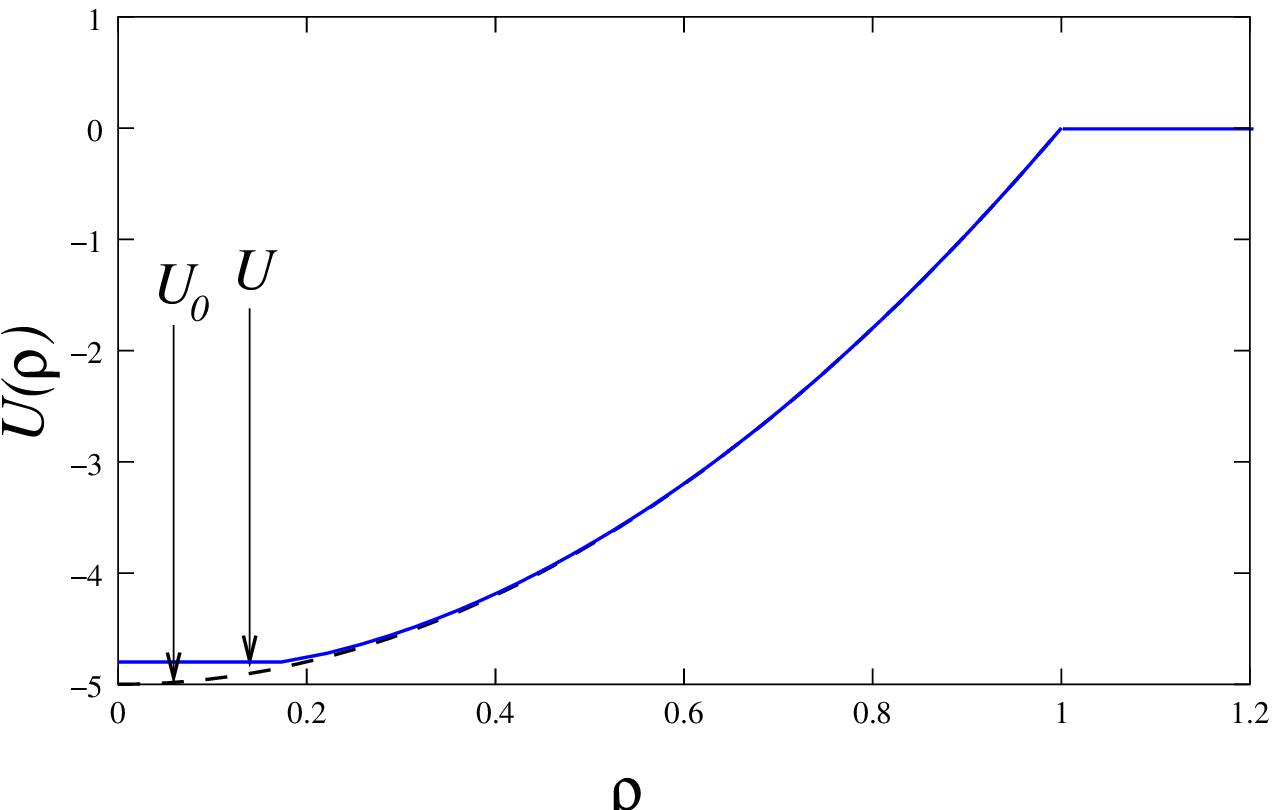}} \par}

\caption{External potential \protect\( {\cal U}_{0}(\rho )\protect \) (dashed
line) and self-consistent potential \protect\( {\cal U}(\rho )\protect \) (full
line) for a super-critical condition \protect\( {\cal M}=-4.8\protect \)
( \protect\( \bar{n}=0.0031n_{c}\protect \) ) with a potential depth
\label{pot-2}\protect\( {\cal V}=-5\protect \), and an interaction constant
\protect\( {\cal W}=0.05\protect \).}
\end{figure}

\begin{figure}
{\centering \resizebox*{12cm}{8cm}{\includegraphics{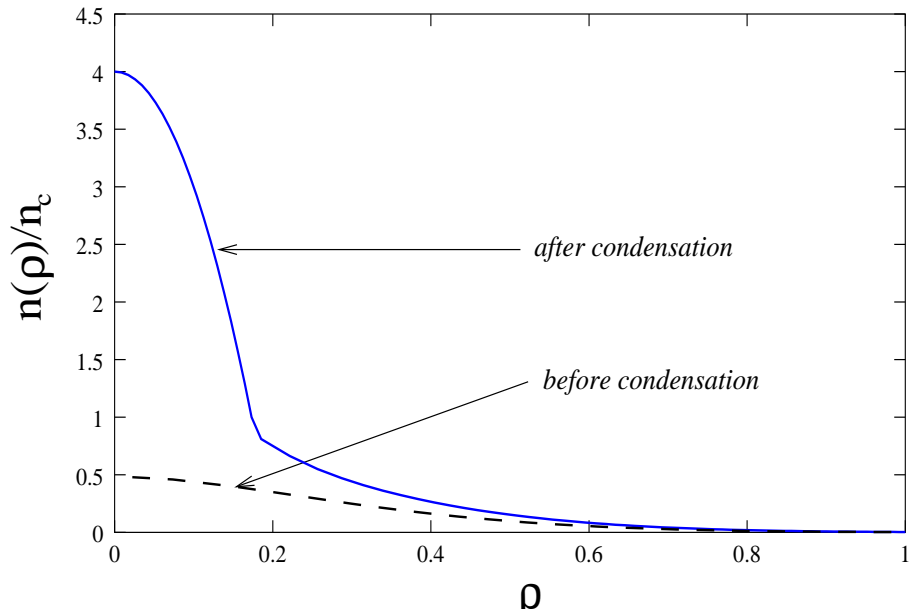}} \par}

\caption{The local density \protect\( n(\rho )/n_{c}\protect \) for the sub-
and super-critical overall densities \protect\( \bar{n}=0.0021n_{c}\protect \) ( dashed line)
and \protect\( \bar{n}=0.0031n_{c}\protect \)  (full line), respectively with
 a potential depth \label{dens-1}\protect\( {\cal V}=-5\protect \), and an
interaction constant \protect\( {\cal W}=0.05\protect \)
.}
\end{figure}

In Fig.\ref{pot-1}, the potentials \( {\cal U}_{0}(\rho ) \) (dotted
line) and \( {\cal U}(\rho ) \) (full line) are shown for a chemical
potential \( {\cal M}=-5.2 \) just below the minimum \( {\cal V} \)
of the external potential, i.e., slightly below the onset of the local
condensation. One sees that the self-consistent potential \( {\cal U}(\rho ) \)
is pushed up. For the chemical potential of \( {\cal M}=-4.8 \) (corresponding to
 a super-critical density) just
above the minimum  \( {\cal V} \) of the external potential, Fig.\ref{pot-2} shows
 that the self-consistent potential
obtains a flat bottom at the chemical potential. This is a typical
signature for the appearance of a condensate here in the range \( 0<\rho <0.2 \).
Fig.\ref{dens-1} shows the resulting peak in the density. It is
remarkable, that the overall densities for the two cases
\( \bar{n}=0.0021n_{c} \) and \( \bar{n}=0.0031n_{c} \) do not differ
very much, both being much lower, than the critical one, while the
condensate density in the second case exceeds the critical density
by a factor \( 4 \) .

\begin{figure}
{\centering \resizebox*{12cm}{8cm}{\includegraphics{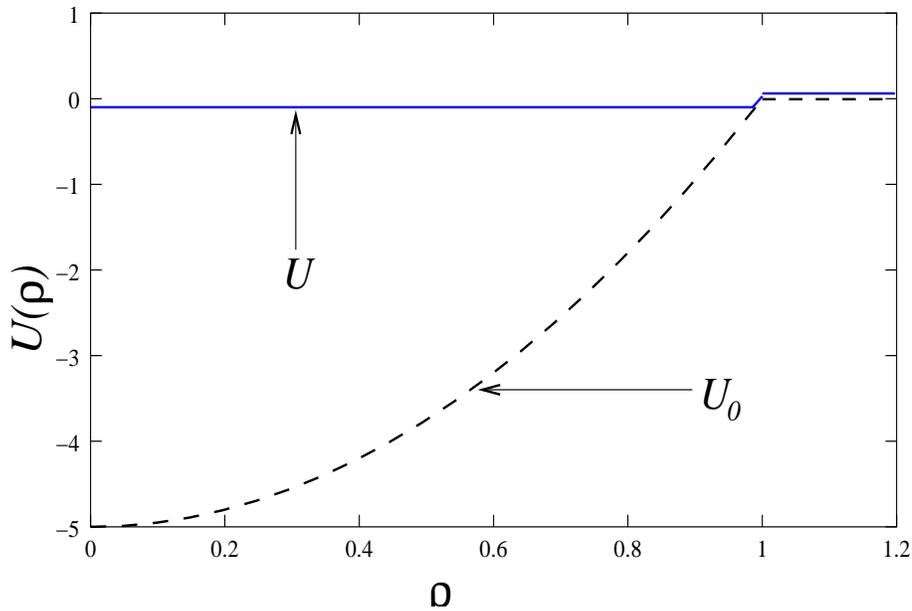}} \par}

\caption{The external potential \protect\( {\cal U}_{0}(\rho )\protect \)
  (dashed line) and the renormalized potential \protect\( {\cal U}(\rho )\protect \)
(full line) for the normalized chemical potential \protect\( {\cal
  M}=-0.1\protect \), and the corresponding density ( \protect\( \bar{n}=0.584n_{c}\protect \)
) with a potential depth \label{pot-3}\protect\( {\cal V}=-5\protect \), and
an interaction constant \protect\( {\cal W}=0.05\protect \). }
\end{figure}

\begin{figure}
{\centering \resizebox*{12cm}{8cm}{\includegraphics{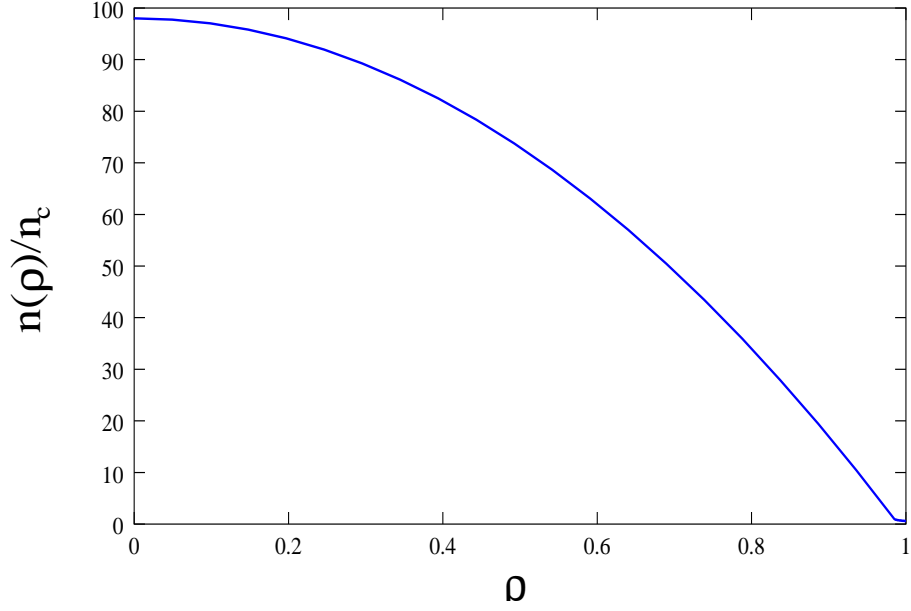}} \par}

\caption{The local density \protect\( n(\rho )/n_{c}\protect \) for the
  chemical potential  \protect\( {\cal M}=-0.1\protect \) (corresponding density
\protect\( \bar{n}=0.584n_{c}\protect \)) with a potential depth
  \label{dens-3} \protect\( {\cal V}=-5\protect \), and an interaction constant
\protect\( {\cal W}=0.05\protect \). }

\end{figure}

A further increase of the chemical potential \( {\cal M} \) (or equivalently of 
the overall density \( \bar{n} \)) will lift the bottom of the self-consistent potential
further as it may be seen on Fig.\ref{pot-3}, while the corresponding
condensate density in the minimum of the self-consistent potential reaches very
high values and the spatial density distribution (shown in Fig.\ref{dens-3})
widens out. Actually it is already very close to \( -\frac{1}{{\cal W}}{\cal U}_{0} \)
, which will be its value at \( \bar{n}=n_{c} \) .

The general feature of the condensation scenario is seen most clearly in
 Fig.\ref{curbe}, where we represented the density peak of the trapped particles and its
half-width  as functions of the overall
 density \(\bar n \) (here with the choice
 \( {\cal{W}} = 0.05\) and \({\cal{V}} = -5 \)). The height 
of the density peak increases with the overall density but its
slope increases abruptly at the overall  density of \(0.0027 n_c \).
The half-width of the local distribution 
at the same overall density has a  minimum resembling to a cusp.  

\begin{figure}
{\centering \resizebox*{12cm}{8cm}{\includegraphics{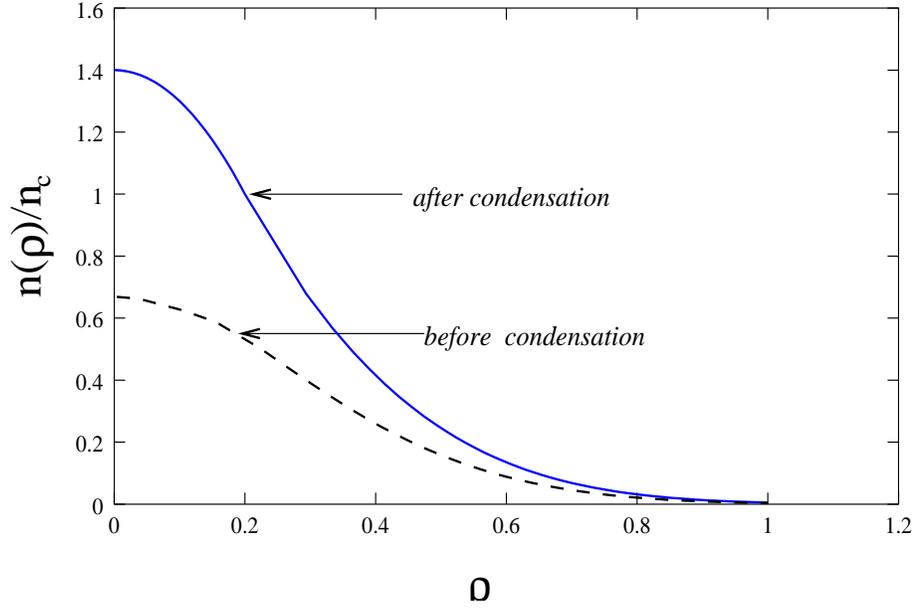}} \par}

\caption{The local density \protect\( n(\rho )/n_{c}\protect \) for the sub-
and super-critical overall densities  at 
\protect\( \bar{n}=0.0035n_{c}\protect \) (dashed line) and  
\protect\( \bar{n}=0.0052n_{c}\protect \) (full line) \label{dens-4} for the
strong interaction case with a potential depth
\protect\( {\cal V}=-5\protect \) , and an interaction constant
\protect\( {\cal W}=0.5\protect \).}
\end{figure}

For a stronger interaction,  corresponding to \( {\cal W}=0.5 \) the
role of the local condensate is not so striking. It can enhance the
density in the well maximally by a factor of 10, resulting in a less
pronounced peak. Nevertheless, the density inside the well may reach
very high values compared to the overall density. This is illustrated
in Fig. \ref{dens-4} for two positions of the chemical potential
\( {\cal M}=-4.7 \) and \( {\cal M}=-4.3 \) corresponding to situations
without and with condensate. In both cases the overall density is
much below the critical one (\( \bar{n}=0.0035n_{c} \) and \( 0.0052n_{c} \)).

\section{Excitons in a local potential}

In recent experiments \cite{Naka2} Naka and Nagasawa have observed
after uniform illumination a strong luminescence in the center of
a strain induced potential well in a 
\( Cu_{2}O \) sample at \( T=2^{\circ }K \). This luminescence may be
 due to the condensation of para-excitons
(having a Bohr radius \( a_{ex}=0.7× 10^{-7}cm \) ). The luminescence
peak was concentrated in a small vicinity around the potential well
minimum. The strain induced potential in their experiment is asymmetric,
but may be approximated qualitatively by a parabolic potential well
like the one we choose having a depth of \( v\approx -1meV \) and
a radius of \( r_{0}\approx 10^{-2}cm \). This would correspond in
our dimensionless parameters to \( {\cal V}=-5 \). In an ideal oscillator
potential with the same curvature one gets an energy spacing of the
quantum mechanical levels of \( \hbar \omega \approx 10^{-4}meV \)
and a very small radial extension of the ground state oscillator wave
function of \( 5× 10^{-5}cm \) for the \( Cu_{2}O \) exciton
mass of \( m=2.7m_{0} \). Since \( v/\hbar \omega \gg 1 \) many
energy levels are contained in the potential well and one may expect
the validity of the quasi-classical description. At this temperature
one has in \( Cu_{2}O \) a critical density of \( n_{c}\approx 10^{17}cm^{-3} \).
The density of the excitons created by laser illumination has not
been measured directly. 

On the other hand, the interaction constant \( w \) is not directly
experimentally accessible. Tentatively one may consider the well-known
potential between hydrogen atoms obtained by the Heitler-London theory
of the hydrogen molecule \cite{Sugiura} and replace the Rydberg energy
and the electron mass, with the corresponding entities of the exciton
(exciton Rydberg \( E_{R} \) and exciton Bohr radius \( a_{B} \)

\begin{eqnarray}
v(r) & = & \frac{E_{R}}{S(r)^{2}}\int d\vec{y}_{1}\int d\vec{y}_{2}\left ( \right. \left ( \phi (\vec{y}_{1})\phi (\vec{y}_{2})± \phi (\vec{y}_{1}+\vec{r})\phi (\vec{y}_{2}-\vec{r})\right) 
\nonumber
\\
& × &   \left( \frac{1}{|\vec{y}_{1}-\vec{y}_{2}+\vec{r}|}+\frac{1}{|\vec{r}|}-\frac{1}{|\vec{y}_{1}+\vec{r}|}-\frac{1}{|\vec{y}_{2}-\vec{r}|}\right) \phi (\vec{y}_{1})\phi (\vec{y}_{2})\left. \right)
\end{eqnarray}
\label{eq:2}

where the normalization factor is

\begin{equation}
S(r)^{2}=1± \int d\vec{y}_{1}\int d\vec{y}_{2}\phi (\vec{y}_{1})\phi (\vec{y}_{1}+\vec{r})\phi (\vec{y}_{2})\phi (\vec{y}_{2}-\vec{r})
\end{equation}

The 1s wave-functions \(\phi(\vec{x}) \) are

\begin{equation}
\phi (\vec{x})=\frac{1}{\sqrt{\pi }a_{B}^{3}}e^{-\frac{|\vec{x}|}{a_{B}}}\: .
\end{equation}

Actually Sugiura \cite{Sugiura} evaluated analytically all the involved
integrals. The positive sign corresponds to the symmetrical state,
while the negative one corresponds to the anti-symmetrical state in
the coordinates. There are also more sophisticated variational calculations
for the bound molecular state (symmetrical state), however not for
the repulsive (anti-symmetrical) case, where a variational approach
is not justified.

\begin{figure}
{\centering \resizebox*{12cm}{8cm}{\includegraphics{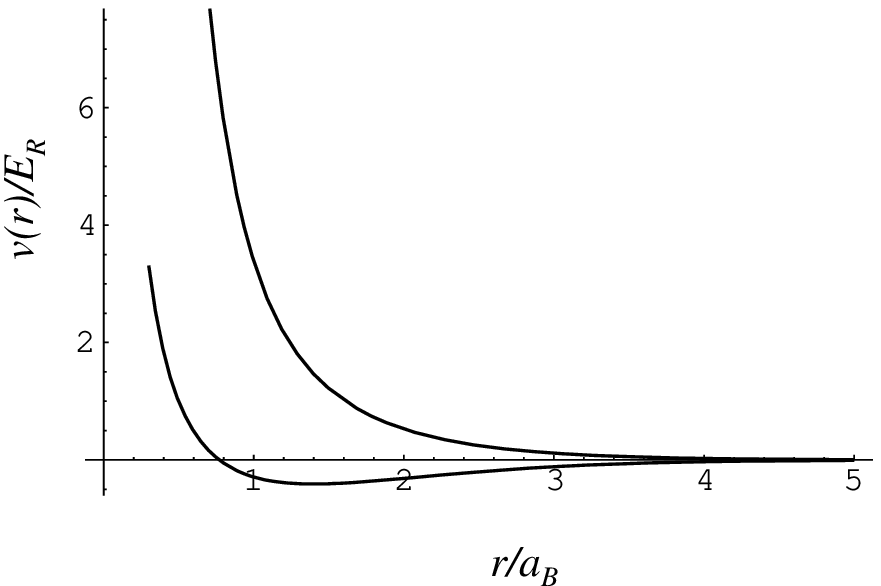}} \par}

\caption{The Heitler-London potential for the symmetrical and anti-symmetrical
configurations (lower respectively upper curve).\label{potential}}
\end{figure}

The para-excitons (singlet spin state) in \(Cu_2 O\)  are the lowest exciton state.
Then antisymmetric wave functions in the coordinates for the identical
particles should be constructed and the resulting force between the
excitons will be repulsive as shown in Fig.\ref{potential}.
The Heitler-London potential decreases with the distance between the
excitons exponentially as \( e^{-\frac{2r}{a_{B}}} \) and the local
approximation should hold. The calculated value of the interaction
constant \( w \) is \( 49.26E_{R}a_{B}^{3} \) respectively. For
\( Cu_{2}O \) this value leads to \( {\cal W}\approx 1 \). This
means that in \(Cu_2 O\)  the strong coupling case is realized. 

\begin{figure}
{\centering \resizebox*{12cm}{8cm}{\includegraphics{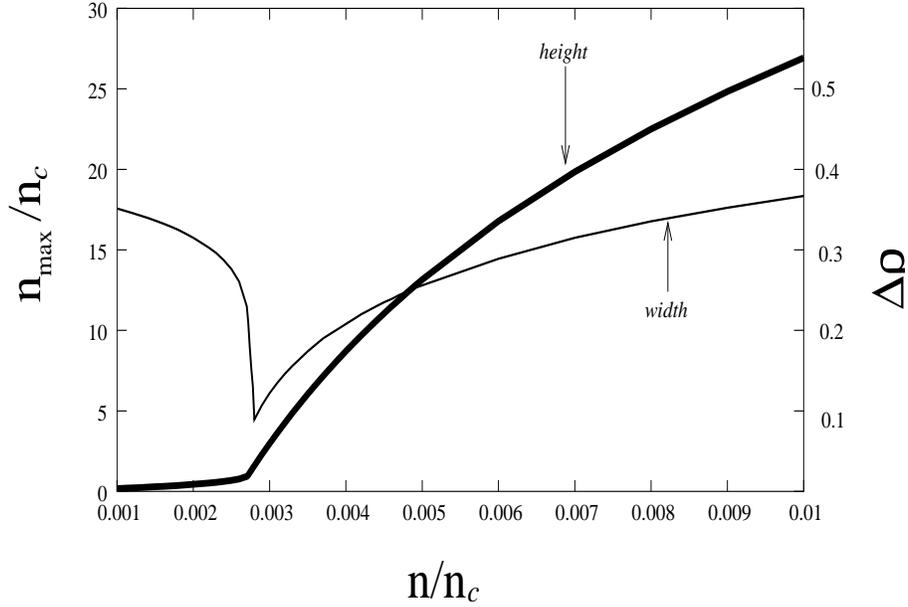}} \par}

\caption{ The dependence of the height (thick line) and halfwidth (thin line)
of the density distribution of the trapped particles
on the overall density in the vicinity of the condensation point. \label{curbe}}
\end{figure}

Because we had to identify  excitons with point like bosons, it is
questionable to consider a Heitler-London potential between the excitons
at distances smaller than \( 2a_{B} \). Actually one should consider
such quantitative estimates with some reservations. Therefore, one cannot
make a reliable quantitative prediction  for the trapping wells in \( Cu_{2}O \).
However, the photoluminescence of the trapped excitons should display the
characteristics of Fig.\ref{curbe} if a local Bose condensation occurs.       

Finally we want to draw the attention to similar observations \cite{Butov}
with excitons confined to an indirect GaAs/AlGaAl quantum well layer.
Here also a strong localized luminescence has been observed due to trapping,
 but the system is rather two-dimensional.

Actually all the considerations relating to Bose-Einstein condensation of excitons
 are subject to the condition, that the electron-hole pairs are mainly bound into
 excitons, which under experimental conditions may be too restrictive.  

In conclusion, we have formulated a quasi-classical condensation kinetics of
interacting bosons which are in contact with a thermostat and trapped in a finite
macroscopic potential well. The theory is based on the coupled equations
 Eqs.\ref{Boltz-cond}-\ref{pot-s.c.} for the non-condensate and condensate.
  The equilibrium solution has been discussed in detail. Due to the repulsive interaction,
 the trapping
potential becomes renormalized after the condensation has occurred, allowing
the chemical potential to increase over the minimum of the well for
an increasing overall particle density. In particular it has been shown
that the spatial
distribution of the trapped particles shows distinct changes as a local Bose
Einstein condensation occurs. The possible relation of these results to recent
experiments with trapped excitons has been discussed.

One of the authors (A.M.B) thanks the Deutsche Forschungsgemeinschaft for
the generous support allowing his stay at the University of Frankfurt.
We appreciate private communications on their experiments by N. Nagasawa
and D.S. Chemla.

\newpage

\begin{thebibliography}{1}
\bibitem{Cornell} M.H. Anderson, J.R. Ensher, M.R. Matthews, C. E. Wiemann
 and E.A. Cornell et al., Science \textbf{269}, 198 (1995)
\bibitem{Ketterle} K.B. Davis, M.O. Mewes, M.R. Andrews, N.J. van Druten,
D.S. Durfee, D. M. Kurn, and W. Ketterle,  Phys. Rev. Lett. \textbf{75}, 3969 (1995)
\bibitem{review}F. Dalfovo, S. Giorgini,  L. P. Pitaevskii, and S. Stringari,
Rev. Mod. Phys. \textbf{71}, 463 (1999)
\bibitem{Pethick} C. J. Pethick, H. Smith, Bose -Einstein Condensation in Dilute Gases,
 Cambridge University Press, Cambridge (2002) 
\bibitem{PitString} L.Pitaevskii, S. Stringari, Bose-Einstein Condensation, Oxford Science Publications, 
 Clarendon Press, Oxford (2003)
\bibitem{Schmitt}A. Schmitt, L. Banyai, and H. Haug, Phys. Rev. \textbf{B62}, 205113-1
(2001)
\bibitem{Naka2}N. Naka and N. Nagasawa, Solid State Comm. \textbf{126}, 523 (2003) 

\bibitem{Kadanoff-Baym}L. P. Kadanoff and G. Baym, Quantum Statistical Mechanics, Benjamin,
New York (1962)
\bibitem{Liboff}R. L. Liboff, Kinetic Theory, Prentice Hall, London, 1990
\bibitem{condboltz}L. Banyai, P. Gartner, O.M. Schmitt, and H. Haug, Phys. Rev. \textbf{B61},
8823 (2000)
\bibitem{condvol}L. Banyai and  P. Gartner, Phys. Rev. Lett. \textbf{88}, 210404-1, (2002)

\bibitem{kinquantbos}O. M. Schmitt, D. B. Tran Thoai, L. Banyai, P. Gartner, and H. Haug,
Phys. Rev. Lett. \textbf{86}, 3839 (2001)
\bibitem{Butov}L. V. Butov, C. W. Lai, A. L. Ivanov, A. C. Gossard, and D. S. Chemla,
Nature \textbf{427}, 47 (2002) and private communications


\bibitem{Sugiura}Y.Sugiura, Z. Physik \textbf{45}, 484 (1927)
\end{thebibliography}
\end{document}